\documentclass[twoside,pra,a4paper,10pt,tightenlines,twocolumn]{revtex4}
\usepackage{amsmath,amssymb}
\newif\ifpdf
	\ifx\pdfoutput\undefined
	\pdffalse 
	\else
	\pdfoutput=1 
	\pdftrue
	\fi
\ifpdf
	\usepackage[pdftex]{graphicx}
	\DeclareGraphicsExtensions{.pdf}
\else
	\usepackage[dvips]{graphicx}
	\DeclareGraphicsExtensions{.eps}
\fi
\begin{document}
\title{Critically-bound four-body molecules}
\author{J.-M.~Richard}
\affiliation{Institut des Sciences Nucl\'eaires, Universit\'e Joseph 
Fourier -- CNRS-IN2P3,
53, avenue des Martyrs, F-38026 Grenoble Cedex, France}
\date{\today}
\begin{abstract}
The $(p,d,\bar{p},\bar{d})$ molecule, with a proton, a deuteron and their antiparticles, is stable againt spontaneous dissociation, but none of its three-body subsystems are stable. This molecule should be built by combining two atoms, for instance a protonium $(p\bar{p})$ and its heavier analogue $(d\bar{d})$. Most other four-body molecules have at least one stable three-body subsystem  and thus  can be built by adding the constituents one by one.
\end{abstract}
\maketitle
Bressanini et al.\ \cite{Bressanini97} have studied the stability of  four-charge systems with masses $(M^+,m^+,M^-,m^-)$. For $M=m$, this corresponds to the positronium molecule $\mathrm{Ps}_2$, whose stability was first demonstrated in 1947 \cite{Hylleraas47a}. For $M\gg m$ or $M\ll m$, this is a hydrogen--antihydrogen $\mathrm{H}\overline{\mathrm H}$ system (without annihilation, strong interaction, etc.) which hardly competes with the deeply bound protonium $(M^+M^-)$ involved in the lowest threshold $(M^+M^-)+(m^+m^-)$.  Stability is thus restricted to an interval of $M/m$ close to unity. The Monte-Carlo calculation of Ref.~\cite{Bressanini97} leads to an estimate
\begin{equation}\label{MmMm}
{1\over 2.2}\lesssim {M\over m}\lesssim2.2~,
\end{equation}
which is confirmed by a powerful variational method \cite{Varga99}.

The case of three unit charges  is well documented \cite{Martin92,Armour93, Armour2003}, in particular for the $(M^\pm,m^\mp,m^\pm)$ configurations. For $M=m$, this is the stable positronium ion Ps$^-$. For $M\gg m$, we have $(p,e^-,e^+)$, and for $M\ll m$, $(\bar{p},p,e^-)$, both unstable. Mitroy \cite{Mitroy2000}, using the same stochastic variational approach as in  Ref.~\cite{Varga99}, found that stability is confined to 
\begin{equation}\label{Mitroy}
0.70\lesssim M/m \lesssim1.64~.
\end{equation}

Comparing the results (\ref{MmMm}) and  (\ref{Mitroy}) indicates a window for ``Borromean'' binding. For instance, for $M/m=2$, which is the deuteron-to-proton mass ratio, the $(M^+,m^+,M^-,m^-)$ molecule  is bound, but neither $(M^\pm,m^\mp,M^\pm)$ nor $(m^\pm,m^\mp,M^\pm)$ are stable.

The word ``Borromean'' has been proposed in nuclear physics to identify bound states whose subsystems are unbound \cite{Bang93}. It comes from the Borromean rings, which are interlaced in such a subtle topological way, that if any one of them is removed, the two others become unlocked. 
For instance, the ${}^6\mathrm{He}$ isotope of ordinary Helium is stable, while ${}^5\mathrm{He}$ is not. In a three-body picture, this means that the ($\alpha, n, n)$ system is bound, whereas $(\alpha, n)$ and $(n,n)$ are unbound.

For $N>3$ constituents, one might define Borromean binding as the property of  
all $N'$-body subsystems being unstable, with $N'=2$, or $N'=N-1$, or  $N'<N$. We propose the following definition:%
{\sl A  bound state is Borromean if there is no path to build the system via a series of  stable bound states by adding the constituents one by one.}
Then, $(p,d,\bar{p},\bar{d})$ is Borromean.  It is truly an atom--atom composite,  more representative of larger molecules of ordinary chemistry. The same is true for neighboring systems $(m_1^+,m_2^+,m_3^-,m_4^-)$ with less symmetry. A minimal extension of the domain of stability can be derived using the variational principle \cite{Richard2002}.

In comparison, $\mathrm{H}_2$ or $\mathrm{Ps}_2$ systems appear to be more robust, with several three-body subsystems being stable, $(p,e^+,e^-)$ or $(p,p,e^-)$ for H$_2$, and $(e^\pm,e^\mp,e^\mp)$ for Ps$_2$. The positronium hydride PsH $(p,e^+,e^-,e^-)$ contains the unstable $(p,e^+,e^-)$, but also the stable $(p,e^-,e^-)$ and $(e^+,e^-,e^-)$, and thus is not Borromean. 

Note that if the antideuteron is replaced by the celebrated $\Omega^-$ hyperon (predicted by Gell-Mann by symmetry considerations which led to the quark model, and discovered by Samios et al. at Brookhaven \cite{Pais:1986nu}), and if the deuteron is replaced by  $\overline{\Omega}{}^+$, the mass ratio $M/m=1.78$ becomes close to one of the critical values of Eq.~(\ref{Mitroy}). If $A=(\Omega^-\overline{\Omega}{}^+)$, we have an effective $(A,p,\bar{p})$ three-body system with both $(A,p)$ and $(A,\bar{p})$ energies vanishing. The Efimov effect \cite{Efimov70} survives finite-size effects, since it is governed by the  long range part of the interaction. However, the Coulomb attraction between $p$ and $\bar{p}$ spoils the $-1/\rho^2$ behavior ($\rho$ is the hyperradius) necessary in the hypercentral potential for Efimov states to appear. See, e.g., the approach by  Fedorov and Jensen, in Ref.~\cite{Efimov70}.

{This investigation grew out of discussions with Dario Bressanini and Andr\'e Martin. Comments by E.A.G.~Armour, A.J.~Cole, A.S.~Jensen and K.~Varga  are also gratefully acknowledged. A partial and preliminary version of this comment was presented at the Few-Body Conference in Bled \cite{Richard2002}.}

\end{document}